\documentclass[11pt]{article}
\usepackage{latexsym}
\usepackage{amssymb}
\usepackage{amsmath}
\usepackage[hypertex]{hyperref}
\usepackage{graphicx}

\newcommand{\lsim}{\lesssim}
\newcommand{\gsim}{\gtrsim}

\begin{document}

\pagestyle{empty}

\begin{flushright}
SLAC-PUB-12189\\
hep-ph/0611111 \\
\end{flushright}

\vspace{2.5cm}

\begin{center}

{\bf\LARGE Gauge mediation in supergravity 
\vspace*{3mm}\\
and gravitino dark matter}
\\

\vspace*{1.5cm}
{\large 
Masahiro Ibe and
Ryuichiro Kitano} \\
\vspace*{0.5cm}

{\it Stanford Linear Accelerator Center, Stanford University,
                Stanford, CA 94309 and} \\
{\it Physics Department, Stanford University, Stanford, CA 94305}\\

\vspace*{0.5cm}

\end{center}

\vspace*{1.0cm}

\begin{abstract}

Gravitinos and hidden sector fields often cause a cosmological disaster
in supersymmetric models. We find that a model with gravitational gauge
mediation solves such a problem quite naturally. The $\mu$-problem is
also absent in the model. Moreover, the abundance of gravitinos explains
correct amount of dark matter of the universe.  The dark matter
abundance can be calculated without detailed information on the thermal
history of the universe such as the reheating temperature after
inflation.

\end{abstract} 

\newpage
\baselineskip 18pt
\setcounter{page}{2}
\pagestyle{plain}
\section{Introduction}

Supersymmetric standard model is advertised as a great successful theory
in particle physics and cosmology. The hierarchy problem is completely
solved by the presence of superpartners. A perturbative nature of the
model allows us to calculate gauge coupling evolutions in high energy
and remarkably the three coupling constants meet at very high
energy. This is a strong indication of supersymmetric grand
unification~\cite{Dimopoulos:1981zb}. In cosmology, the neutralino, the
superpartner of the gauge or Higgs fields, can account for dark matter
of the universe. The thermal relic abundance of the neutralino is in the
right ballpark for masses of order 100~GeV.

However, when we look at actual models of supersymmetry breaking and its
mediation, we encounter many unnatural aspects or even inconsistencies.
In gravity mediated supersymmetry breaking
scenario~\cite{Chamseddine:1982jx}, there is a serious cosmological
difficulty~\cite{Coughlan:1983ci,Banks:1993en} as well as problems of
flavor changing neutral current (FCNC) and CP violating processes. A
supersymmetry breaking sector necessarily requires to contain a gauge
singlet field whose vacuum expectation value in the $F$-component gives
masses to gauginos. After inflation, the scalar component of the singlet
field starts coherent oscillation.
Once it dominates over the energy density of the universe, it either
spoils the success of Big-Bang Nucleosynthesis
(BBN)~\cite{Coughlan:1983ci} or overproduces
gravitinos~\cite{Ibe:2006am} (see~\cite{Dine:1983ys,Joichi:1994ce} for
earlier works).
Therefore, for the mechanism of thermal relic neutralinos as dark matter
to work, we need extra assumptions on the cosmological history so that
we can avoid the oscillation to dominate the energy density.
Models with anomaly mediation~\cite{Randall:1998uk,Giudice:1998xp} can
prevent such oscillation since singlet fields are not necessary to
generate the gaugino masses. However, the annihilation cross section of
the lightest neutralino, the Wino in this scenario, is too large for the
thermal relic to naturally explain the dark matter abundance.
There seems to be no coherent picture for cosmology of supersymmetric
models.

Models of gauge mediation~\cite{Dine:1981za,Dine:1993yw,Dine:1994vc}
also suffer from serious difficulties.
The neutralino cannot be a stable particle since the gravitino is always
the lightest supersymmetric particle (LSP) in this scenario. Moreover,
enhanced production cross sections of gravitinos bring up a danger of
overproduction in the thermal
bath~\cite{Moroi:1993mb,deGouvea:1997tn,Bolz:2000fu,Pradler:2006qh}. This
restricts the reheating temperature of the universe to be low enough,
which makes baryogenesis difficult.
There is also a naturalness issue in the electroweak sector, i.e., the
$\mu$-problem. We need to give a supersymmetric mass term, $\mu$-term,
for the Higgs fields which needs to be similar size to supersymmetry
breaking terms. Unlike in gravity mediation models, we cannot relate
this $\mu$-term to the gravitino mass via the Giudice-Masiero
mechanism~\cite{Giudice:1988yz} because the gravitino mass is too
small. It is possible to assume a direct coupling between the Higgs
sector and a supersymmetry breaking sector to obtain the $\mu$-term, but
it often induces too large $B$-term.~\cite{Dine:1994vc}.

We cannot say that the minimal supersymmetric standard model (MSSM) is a
successful model without having a consistent picture for supersymmetry
breaking/mediation. 
It has been more than twenty years since discussions of supersymmetry
phenomenology started, and until recently it has served as the most
promising candidate of physics beyond the standard model.
However, recent various progress in astrophysical observations and
theoretical calculations such as the measurement of the matter density
of the universe~\cite{Spergel:2003cb} and BBN constraints on the
gravitino abundance~\cite{Kawasaki:2004qu} put a threat to many
supersymmetric models.
At the same time, these are new strong hints which guide us to the true
theory.
In this paper, we find that a simple model of gauge mediation actually
provides us with a viable and natural scenario for cosmology and
phenomenology.

The model we discuss consists of a singlet field $S$ and messenger
fields $q$ and $\bar{q}$. The singlet field $S$ breaks supersymmetry
spontaneously and that is mediated to the MSSM sector by loop diagrams
of the messenger fields.
The supersymmetry breaking vacuum is stabilized by gravitational
effects~\cite{Kitano:2006wz}.
This is the entire model.
Under reasonable assumptions, the $S$ field starts coherent oscillation
in the early universe. If we ignore interactions of $S$ with the
messenger fields, decays of the $S$ field either destroy BBN or
overproduce gravitinos, which is the same problem in gravity mediation
scenarios. A renormalizable coupling to the messengers, however,
drastically modifies the decay property of $S$.  Especially, $S$ can
decay into the MSSM particles well before BBN, and, moreover, the
branching fraction to the gravitinos is significantly suppressed.
With gaugino masses of $O(100)$~GeV, gravitinos produced by the $S$
decay can be dark matter of the universe when the gravitino mass is
$m_{3/2} = O(10)$~MeV or $O(500)$~MeV, which corresponds to the mass of
$S$ to be $m_S = O(100)$~GeV and $O(400)$~GeV, respectively.
In fact, this is the most interesting parameter range from the
consideration of the $\mu$-problem.
The $\mu$-problem can be solved by the mechanism of $\mu$-term driven
supersymmetry breaking scenario~\cite{Kitano:2006wm} when $m_S \sim
O(\mu)$.

In the next section, we explain the gauge-mediation model we
consider. The concept of $\mu$-term driven supersymmetry breaking is
also discussed. In Section~\ref{sec:evolution}, we discuss cosmological
evolutions of the $S$ field. We show that $S$ rolls into a preferable
(supersymmetry breaking) vacuum in generic situations.
The decay property of $S$ is discussed in
Section~\ref{sec:dark-matter}. The BBN constraint on parameters and the
abundance of non-thermally generated gravitino dark matter are
calculated.
The discussion on the initial condition for the $S$ oscillation is given
in Appendix~\ref{app:amplitude}.

\section{A simple model of gauge mediation}

\subsection{Gravitational gauge mediation}
The model we study has a very simple structure.
We simply introduce messenger fields to any supersymmetry breaking model
of the O'Raifeartaigh type, i.e., models in which the superpotential
contains a linear term of a gauge-singlet chiral superfield.
The low-energy effective Lagrangian is defined by the following K{\"
a}hler potential and superpotential:
\begin{eqnarray}
  K = q^\dagger q + \bar{q}^\dagger \bar q
+ S^\dagger S - \frac{(S^\dagger S)^2}{\Lambda^2} + \cdots
\end{eqnarray}
\begin{eqnarray}
 W = m^2 S - \lambda S q \bar{q} + c \ ,
\label{eq:superpotential}
\end{eqnarray}
where $S$ is a gauge singlet field in the O'Raifeartaigh model and $q$
and $\bar{q}$ are the messenger fields which have quantum numbers of the
standard model gauge group.
We introduced an interaction term $\lambda S q \bar{q}$ where $\lambda$
is a coupling constant.
The constant term $c$ contributes to the cosmological constant through
$-3|W|^2$ term in a scalar potential.
The non-renormalizable term $-(S^\dagger S)^2/\Lambda^2$ in the K{\"
a}hler potential represents the effect of the massive fields (in the
O'Raifeartaigh model) which have been integrated out.
Although this model does not break supersymmetry in the limit of the
Planck scale $M_{\rm Pl}$ to infinity, a supersymmetry breaking vacuum
appears far away from the origin of $S$ after taking into account
supergravity effects~\cite{Kitano:2006wz}. At the supersymmetry breaking
vacuum,
\begin{eqnarray}
 \langle S \rangle = \frac{ \sqrt 3 \Lambda^2 }{ 6 M_{\rm Pl} }\ ,\ \ \ 
 \langle q \rangle = \langle \bar q \rangle = 0 \ ,
\label{eq:vev}
\end{eqnarray}
and 
\begin{eqnarray}
 F_S = m^2\ .
\end{eqnarray}
The messenger particles obtain masses, $m_q = \lambda \langle S
\rangle$, in this vacuum.
This vacuum is meta-stable when
\begin{eqnarray}
\frac{m^2 M_{\rm Pl}^2}{\Lambda^4} \ll \lambda \ll
\frac{4 \pi}{\sqrt N_q} \frac{\Lambda}{M_{\rm Pl}}\ ,
\end{eqnarray}
where $N_q$ is the number of components in $q$, e.g., $N_q = 5$ for a
pair of $q$ and $\bar q$ in ${\bf 5}$ and ${\bf \bar 5}$ representation
of SU(5).

The constant term $c$ in the superpotential is chosen so that the
cosmological constant is cancelled at the supersymmetry breaking vacuum,
i.e.,
\begin{eqnarray}
 c = m_{3/2} M_{\rm Pl}^2\ ,
\end{eqnarray}
where the gravitino mass is given by
\begin{eqnarray}
 m_{3/2} = \frac{m^2}{\sqrt 3 M_{\rm Pl}}\ .
\end{eqnarray}
The scalar component of $S$ obtains a mass by supersymmetry breaking
effects:
\begin{eqnarray}
 m_S = \frac{2 m^2}{\Lambda}\ .
\label{eq:mS}
\end{eqnarray}

Supersymmetry breaking by $F_S = m^2$ induces gaugino masses by one-loop
diagrams with messengers $q$ and $\bar q$. It is calculated to be:
\begin{eqnarray}
 m_\lambda = \frac{g^2 N}{(4 \pi)^2} \frac{F_S}{\langle S \rangle}
= \frac{g^2 N}{(4 \pi)^2}
\frac{2 \sqrt 3 m^2 M_{\rm Pl}}{\Lambda^2}\ ,
\label{eq:gaugino}
\end{eqnarray}
where $N$ is the number of the messenger particles ($N=1$ for a pair of {\bf 5}
and ${\bf \bar 5}$ representation of SU(5)). 
The phenomenological requirement that the gaugino mass should be
$O(100)$~GeV determines the ratio $m/\Lambda$.

Among three parameters $m^2$, $\Lambda$ and $\lambda$ in this model, low
energy physics is not very sensitive to $\lambda$.  It only appears
through $\log m_q$ where $m_q \equiv \lambda \langle S
\rangle$ is the messenger scale of gauge mediation.
With fixed gaugino masses, we are left with a single parameter $\Lambda$.

\subsection{{\boldmath $\mu$}-term driven supersymmetry breaking}
There is a particularly interesting scale for $\Lambda$ from the consideration
of UV completions of the theory. Among various possible UV completions
(see Ref.~\cite{Kitano:2006wz} for various possibilities), the simplest
one is to identify the field $S$ with a meson field of a strongly
coupled gauge theory.
For example, in an SO(5) gauge theory with a quark $T$ and a
superpotential term:
\begin{eqnarray}
 W = \mu T T\ ,
\label{eq:mass}
\end{eqnarray}
the low energy effective theory below the dynamical scale $\overline
\Lambda$ is simply
\begin{eqnarray}
 W = \mu \overline \Lambda S\ ,
\end{eqnarray}
in one of two branches of the theory. Supersymmetry is broken by $F_S =
\mu \overline \Lambda$. This vacuum is found to be meta-stable recently
by Intriligator, Seiberg and Shih~\cite{Intriligator:2006dd}.
The K{\" a}hler potential is unknown but can be parametrized to be
\begin{eqnarray}
 K = S^\dagger S -  \frac{\eta (S^\dagger S)^2}{\overline{\Lambda}^2} + \cdots \ .
\end{eqnarray}
If $\eta$ is positive, this is exactly the same effective theory we
discussed in the previous subsection.  The scale $\Lambda$ is identified
with $\overline \Lambda / \eta^{1/2}$.
The messenger fields couple to $S$ through $W \ni T T q \bar q / M_*$ in
the superpotential with $1/M_*$ the coefficient of the operator. The
$\lambda$ parameter in Eq.~(\ref{eq:superpotential}) is then $\lambda =
\overline \Lambda/M_*$.

In this model, the $\mu$-term in Eq.~(\ref{eq:mass}) can actually be the
$\mu$-term for the Higgs fields $W \ni \mu H_u H_d$ in models with
composite Higgs fields~\cite{Kitano:2006wm}. The Higgs fields $H_u$ and
$H_d$ are meson fields $H_u \sim (QT)$ and $H_d \sim (\bar Q T)$ where
$Q$ and $\bar Q$ carry the standard model quantum numbers. The mass term
for $T$ in Eq.~(\ref{eq:mass}) therefore induces a mass term for
mesons. (For detailed discussion, see~\cite{Kitano:2006wm}.)
Since $S$ and the Higgs fields $H_u$ and $H_d$ are strongly coupled, we
expect to have terms in the K{\" a}hler potential:
\begin{eqnarray}
 K \ni 
\frac{1}{\overline{\Lambda}^2} S^\dagger S H_u^\dagger H_u
+ \frac{1}{\overline{\Lambda}^2} S^\dagger S H_d^\dagger H_d \ .
\label{eq:higgs-kahler}
\end{eqnarray}
With $F_S = \mu \overline \Lambda$, these terms induce the soft supersymmetry
breaking terms
\begin{eqnarray}
 m_{H_u}^2 \sim m_{H_d}^2 \sim \mu^2\ ,
\end{eqnarray}
independent of the dynamical scale $\overline \Lambda$. This provides a solution
to the $\mu$-problem in gauge mediation.
The $\mu$-term is related to the supersymmetry breaking parameters
because it is the source of the supersymmetry breaking.

A combination of the gravitational gauge mediation and $\mu$-term driven
supersymmetry breaking provides a completely natural scenario for gauge
mediation. The $m^2$ parameter in Eq.~(\ref{eq:superpotential}) is
$O(\mu \Lambda)$ in this scenario (under a reasonable assumption that
$\eta = O(1)$), and it fixes the dynamical scale $\overline \Lambda
(\sim \Lambda )$ by the requirement $\mu \sim m_\lambda \sim O(100)$~GeV
from Eq.~(\ref{eq:gaugino}). For having this relation, the factor
$M_{\rm Pl}/\Lambda$ needs to compensate the one-loop factor $g^2 N / (4
\pi)^2$, which determines the dynamical scale to be of the order of the
GUT scale $\Lambda \sim 10^{16}$~GeV .
This is indicating a close connection between dynamics of supersymmetry
breaking and spontaneous symmetry breaking in grand unified theories.
In fact, the explicit model in Ref.~\cite{Kitano:2006wm} is based on a
GUT model.
A meson field made of $Q$ and $\bar Q$ introduced before plays a role of
the Higgs field which breaks GUT symmetry dynamically.
This mechanism simultaneously provides us with a solution to the
doublet-triplet splitting problem with sufficiently suppressed
dimension five operators for proton decays.

With $m^2 \sim \mu \Lambda$, the mass of scalar component of $S$, $m_S$,
is $O(\mu) \sim O(100-1000)$~GeV (see Eq.~(\ref{eq:mS})), and the
gravitino mass is $O(0.01-1)$~GeV.
We will use $m_S$ and $m_{3/2}$ as free parameters instead of $m^2$ and
$\Lambda$. Relations among them are:
\begin{eqnarray}
 m^2 = ( 1 \times 10^9~{\rm GeV} )^2
\times
\left(\frac{m_{3/2}}{500~{\rm MeV}}\right)\ ,
\end{eqnarray}
\begin{eqnarray}
 \Lambda = 1 \times 10^{16}~{\rm GeV}
\times
\left(\frac{m_S}{400~{\rm GeV}}\right)^{-1}
\left(\frac{m_{3/2}}{500~{\rm MeV}}\right)\ .
\end{eqnarray}
The Bino mass is expressed in terms of these two by
\begin{eqnarray}
 m_{\tilde B} =
200~{\rm GeV} 
\times N
\left(\frac{m_S}{400~{\rm GeV}}\right)^2
\left(\frac{m_{3/2}}{500~{\rm MeV}}\right)^{-1}\ .
\label{eq:bino}
\end{eqnarray}
There are no $O(1)$ ambiguities in these relations.

\section{Cosmological evolution of {\boldmath $S$}}
\label{sec:evolution}

As we have seen in the previous section, the $S$ field is stabilized at
the supersymmetry breaking vacuum by supergravity effects.
However, there is a true and unwanted supersymmetric vacuum at $S = 0$
and $q = \bar{q} = \sqrt{m^{2}/\lambda} $.  It is thus non-trivial
whether $S$ can settle to the supersymmetry breaking minimum in the
thermal history of the universe. This question is addressed in
Ref.~\cite{jay} where it is concluded that thermal effects make the
supersymmetric vacuum more attractive once the messenger particles are
thermalized. In our scenario, it is possible that the temperature of the
universe never exceeds the mass of messenger particles so that thermal
effects are not important.

We show in this section that $S$ runs into the supersymmetry breaking
minimum under reasonable assumptions. It is found to be rather difficult
to fall into the supersymmetric minimum $S=0$ because it can be thought
of a small hole in the vast $S$ space.

The relevant scalar potential of the $S$ field in the $|S| \lsim
\Lambda$ region is given by
\begin{eqnarray}
\label{eq:scalarpot}
V(S) = m^{4} \left( \frac{4}{\Lambda^{2}} |S|^{2} 
+ \frac{ \lambda^{2} N_q }{16\pi^{2}} \log\left(\frac{|S|^{2}}{\Lambda^{2}} \right) 
+\cdots\right)
- \left( 2 \sqrt{3}m_{3/2}^{2} M_{\rm Pl} S + {\rm h.c.} \right).
\end{eqnarray}
The potential for $|S| > \Lambda$ region depends on a UV completion, but
it is not important for our discussion.
The logarithmic term is induced at one-loop level due to the interaction
with messenger particles.  The supersymmetry breaking vacuum given in
Eq.~(\ref{eq:vev}) is determined by a balance between the mass term and
the linear term.
The logarithmic term is important near $S=0$.
The $S$ field, however, feels much stronger force toward the
supersymmetry breaking vacuum at most points on the complex $S$ plane
since it is supported by a quadratic potential.
This is why the supersymmetry breaking vacuum is more attractive at a
low temperature. If the value of $S$ is not at the origin after
inflation, $S$ rolls down to the $S \neq 0$ vacuum.

The minimum of the scalar potential during inflation is not at $S = 0$
or $S = \langle S\rangle$, but is likely to be around $|S|\sim \Lambda$
(see Appendix~\ref{app:amplitude} for detailed discussion).
Once inflation starts, $S$ is trapped at the minimum $S_0 = O(\Lambda)$
and freezes until the end of inflation.
After inflation, the value of $S$ remains fixed at $ S \simeq S_{0}$ for
a while until the mass term $m_S$ becomes important compared to the
Hubble mass term of $O(H)$, and eventually, it starts to oscillate
about the supersymmetry breaking minimum.
In general, the initial value $S_0$ is not aligned to the $\langle S
\rangle$ direction, which makes motion of $S$ almost confined in a
straight line on the complex plane. Therefore, $S$ never approaches to
the supersymmetric minimum.
The amplitude of the oscillation gradually decays due to the cosmic
expansion and $S$ settles to the supersymmetry breaking minimum.

\begin{figure}[t]
 \begin{center}
  \includegraphics[height=4.8cm]{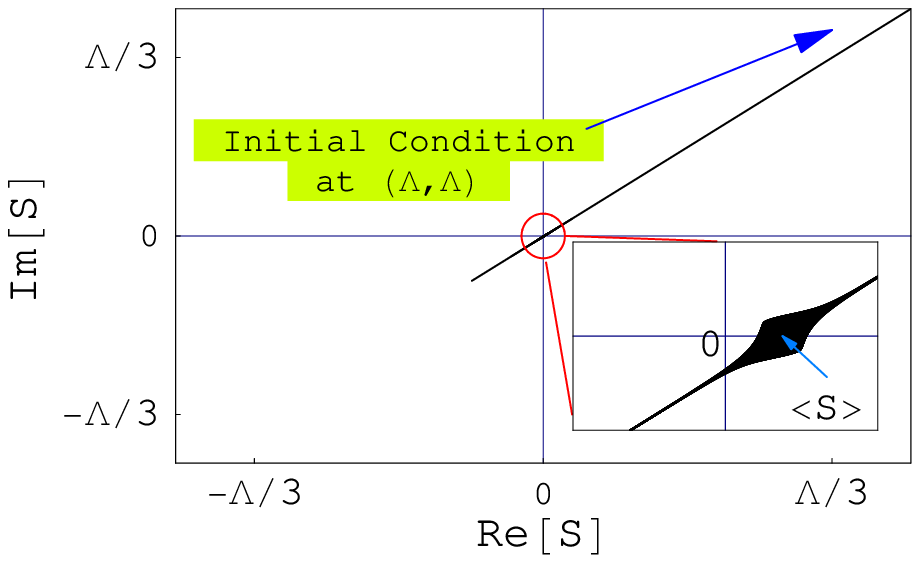}
  \includegraphics[height=4.8cm]{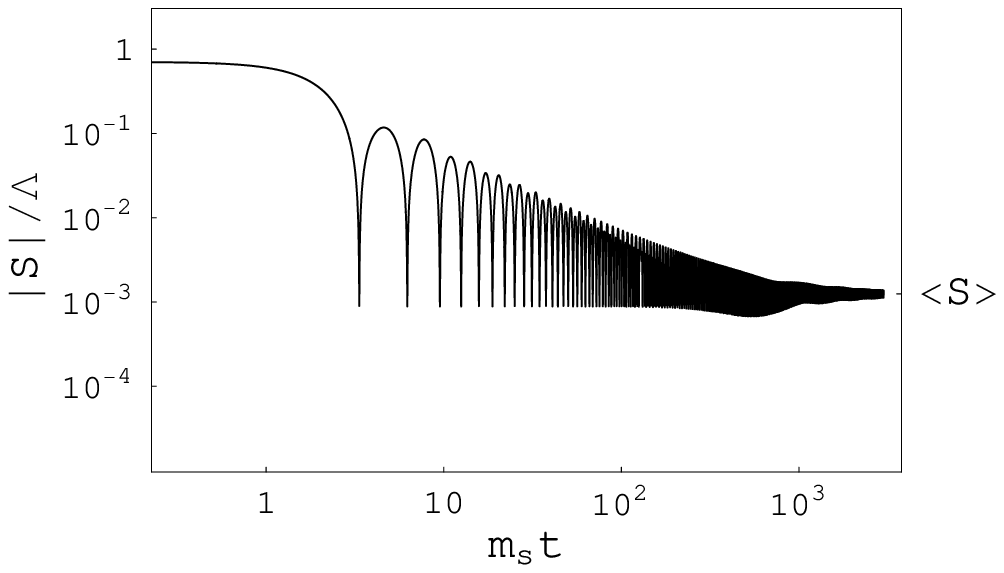}
\end{center}
\caption{A typical evolution of the $S$ field (left) and time dependence
of the amplitude $|S|$ (right) for $m_{S} = 400$\,GeV,
$m_{3/2}=500$\,MeV and $\lambda = 10^{-3}$.
} 
\label{fig:oscillation}
\end{figure}

In Fig.~\ref{fig:oscillation}, we show a typical evolution of the $S$
field for $m_{S} = 400$\,GeV, $m_{3/2}=500$\,MeV and $\lambda =
10^{-3}$.  Here, the initial condition we took is ${\rm Re}(S)= {\rm
Im}(S)= \Lambda$.  As we expected $S$ oscillates along an almost
straight line about the supersymmetry breaking minimum (see also a
closeup in the left panel).  We also show time dependence of the
amplitude $|S|$ during oscillation in the right panel.  The figure shows
that the impact parameter to $S = 0$ during the oscillation is of ${\cal
O}(\langle S\rangle)$ and the amplitude of the oscillation decays
gradually.  Eventually, $S$ settles to the supersymmetry breaking
minimum.

In the above analysis, effects of the logarithmic potential in
Eq.~(\ref{eq:scalarpot}) are not important as long as $\lambda \lsim
10^{-(2-3)}$.
It does attract $S$ to the supersymmetric minimum, but the attractive
force is only important when $S$ approaches very close to the origin.
With an impact parameter of $O( \langle S \rangle )$, the effect is
negligible.

The messenger fields have masses, $m_q = \lambda S$, when $S$ is away
from the origin.
With the large impact parameter of the oscillation of $S$, $q$ and $\bar
q$ are always stabilized at $q = \bar{q} = 0$.

Although they are always massive, time variation of the masses by the
$S$ oscillation causes a non-perturbative production of the messenger
fields~\cite{Kofman:1997yn}. Once they are copiously produced, it gives
rise to a force to attract $S$ to origin so that the energy density is
reduced by minimizing the mass of $q$~\cite{Kofman:2004yc}\footnote{We
thank J.~Wacker for discussion on this point.}. However, since the
variation of $m_q$ is quite adiabatic (i.e., $\dot{m}_q \sim m_S m_q \ll
m_q^2$), the effect is exponentially suppressed.

We need to check here that the messenger particles do not enter thermal
equilibrium for the above discussion to hold.
If they are produced in the thermal bath, it induces a thermal mass term
to $S$ which attracts to $S=0$~\cite{jay}.
However, as long as temperature is always (including the inflaton
oscillation era) lower than $m_q = \lambda S$, production of messenger
particles is also exponentially suppressed.
With the large impact parameter $(O(\langle S \rangle))$ and initial
amplitude $(O(\Lambda))$ of this scenario, the condition can be
satisfied in most inflation models.

\section{Gravitino dark matter}
\label{sec:dark-matter}

We now follow the cosmological evolution after the $S$ oscillation.
A brief summary of the history is the following.
Because of a long lifetime of the $S$ field, its oscillation energy
density eventually dominates over the universe unless the reheating
temperature after inflation is very low. 
Decays of $S$ reheat the universe to a temperature of $O(1-100~{\rm
MeV})$.
The large entropy production from the $S$ decay dilutes preexisting
radiation and matter, including thermally produced superparticles,
gravitinos and also even baryon asymmetry of the universe.
The main decay channel of the $S$ field is into two Higgs bosons ($S \to
h h$) or two gluons ($S \to gg$) depending on $m_S$.
Radiation produced by the $S$ decay takes over the energy density of the
universe below this decay temperature and therefore standard BBN can
happen after temperature drops down to $O({\rm MeV})$.
Decays of $S$ produce gravitinos through a rare decay mode $S \to
\psi_{3/2} \psi_{3/2}$. The produced amount is remarkably consistent
with the observed dark matter density of the universe in the parameter
regions of our interest.

We start discussion with deriving the interaction Lagrangian of $S$,
with which we can calculate the decay width of $S$ and the branching
ratio of the gravitino mode. We will then discuss the gravitino abundance
of the universe.

\subsection{Interaction Lagrangian of the {\boldmath $S$} field}

%
The $S$ field couples to the MSSM particles only through loop diagrams
with messenger particles $q$ and $\bar q$. The effect is encoded in the
$S$ dependence of low energy parameters.

The low energy values of the gauge coupling constants depend on $S$
through $\log S$ since it changes the scale at which $q$ and $\bar q$
are integrated out. The kinetic term of the gauge bosons are
\begin{eqnarray}
 {\cal L}_{\rm kin} = - \frac{1}{8 g^2(S)} F_{\mu \nu} F^{\mu \nu}
+ {\rm h.c.}\ ,
\label{eq:gauge}
\end{eqnarray}
where 
\begin{eqnarray}
 \frac{1}{g^2 (S)} = \frac{1}{g^2_0} 
- \frac{2 (b_L + N)}{(4 \pi)^2} \log \frac{\lambda S}{\Lambda_H}
- \frac{2 b_L}{(4 \pi)^2} \log \frac{\mu_R}{\lambda S}\ ,
\end{eqnarray}
with $g_0$ being the gauge coupling constant at a scale $\Lambda_H$.
The factor $b_L$ is the beta-function coefficient below the messenger
scale and $\mu_R$ is the renormalization scale.
After canonically normalizing the kinetic terms of gauge bosons, we
obtain the interaction terms from Eq.~(\ref{eq:gauge}):
\begin{eqnarray}
 {\cal L}_{F} = 
\frac{g^2 N}{(4 \pi)^2} \frac{1}{\langle S \rangle} 
\cdot \frac{1}{4} S F_{\mu \nu} F^{\mu \nu} + {\rm h.c.}
\label{eq:gauge-int}
\end{eqnarray}

The interaction terms with gauginos can be obtained in a similar
way. From the $S$ dependence of the gaugino mass terms:
\begin{eqnarray}
 {\cal L}_{\rm gaugino} = - \frac{1}{2}
m_\lambda (S) \lambda \lambda + {\rm h.c.}\ ,
\end{eqnarray}
with
\begin{eqnarray}
 m_\lambda (S) = \frac{g^2 N}{(4 \pi)^2} \frac{m^2}{S}\ ,
\end{eqnarray}
we obtain
\begin{eqnarray}
 {\cal L}_{\lambda} =
\frac{1}{2} \frac{m_\lambda}{\langle S \rangle}  S \lambda \lambda
+{\rm h.c.}
\end{eqnarray}
Comparing with Eq.~(\ref{eq:gauge-int}), this gives larger contribution
to the decay width of $S$ by a one-loop factor when $m_S \sim m_\lambda$.

The same conclusion can be obtained for scalar fields. The soft mass
terms are
\begin{eqnarray}
 {\cal L}_{\rm scalar} = 
- m_{\tilde f}^2 (S) \tilde{f}^\dagger \tilde  f\ ,
\end{eqnarray}
with
\begin{eqnarray}
 m_{\tilde f}^2 (S) = 
\left[ \frac{g^2}{(4 \pi)^2} \right]^2
\cdot 2 C_2 N 
\left| \frac{m^2}{S} \right|^2\ .
\label{eq:scalar-mass}
\end{eqnarray}
$C_2$ is the quadratic Casimir factor.
We can read off interaction terms as follows:
\begin{eqnarray}
 {\cal L}_{\tilde f} = 
\frac{m_{\tilde f}^2}{\langle S \rangle}
 S \tilde{f}^\dagger \tilde  f + {\rm h.c.}
\label{eq:scalar}
\end{eqnarray}
Again, this gives larger contribution by a one-loop factor compared to
Eq.~(\ref{eq:gauge-int}). For the Higgs fields, the scalar masses have
two origins, i.e., one from the gauge mediation and the other from
Eq.~(\ref{eq:higgs-kahler}). For the calculation of couplings between
$S$ and the Higgs fields, we should use the gauge-mediation
contribution, i.e., Eq.~(\ref{eq:scalar-mass}), for $m_{\tilde f}^2$ in
Eq.~(\ref{eq:scalar}).
It is amusing that the $S$ field dominantly decays into supersymmetric
particles and the Higgs fields if it is kinematically allowed.

Coupling to gravitinos can be obtained by simply looking at the
Lagrangian. The largest contribution for the $S$ decay comes from the
coupling to the longitudinal mode of the gravitino (the fermionic
component of $S$, $\tilde s$). From the $- (S^\dagger S)^2 / \Lambda^2$
term in the K{\" a}hler potential, we obtain $S^\dagger $ to $\tilde s
\tilde s$ coupling such as
\begin{eqnarray}
 {\cal L}_{3/2} = - \frac{2 F_S^\dagger}{\Lambda^2} S^\dagger \tilde s \tilde s 
+ {\rm h.c.}
\to - \frac{1}{2} \frac{m_{3/2}}{\langle S \rangle} S^\dagger 
\bar{\psi}_{3/2} \psi_{3/2}
+ {\rm h.c.}
\label{eq:gravitino}
\end{eqnarray}
Therefore, the decay width is suppressed by $O((m_{3/2}/m_\lambda)^2)$
compared to the gaugino/scalar modes.
Also, there is a suppression of $O(m_{3/2}/m_S)$ compared to the gauge
boson mode.
This effect is important for the gravitino abundance.

\subsection{{\boldmath $S$} decays and gravitino production}

In the following we consider two cases (A): $m_S > 2 m_h$, where $S$ can
decay into two Higgs bosons and (B): $m_S < 2 m_h$, where $S$ dominantly
decays into two gluons. We find that in both cases gravitinos from the
$S$ decay in early time naturally explain the dark matter component of
the universe.

\vspace{5mm}
\noindent {\bf Case (A)}: The $S \to hh$ decay is open ($m_S > 2 m_h$).\\ 
We further assume here that $S$ dominantly decays into two Higgs bosons.
This assumption is valid as long as $m_S/2$ is smaller than
other SU(2)$_L$ or SU(3)$_C$ charged superparticle masses (except for
the Higgsino). This will be justified later.
 From the experimental lower bound on the Higgs boson mass $m_h \geq
 114$~GeV~\cite{Barate:2003sz}, we obtain $m_S \gtrsim 230$~GeV.
The decay width of $S$ in this case is given by
\begin{eqnarray}
 \Gamma_H = \frac{x_H^2 N^2}{1536 \pi} \frac{m_S^3}{M_{\rm Pl}^2}
\left( \frac{m_S}{m_{3/2}} \right)^8\ ,
\end{eqnarray}
where $x_H$ is a two-loop factor:
\begin{eqnarray}
 x_H = 
\frac{g_2^4}{(4 \pi)^4} \cdot \frac{3}{4}
+\frac{g_Y^4}{(4 \pi)^4} \cdot \frac{5}{3} \cdot \frac{1}{4}
\simeq 6 \times 10^{-6}
\ .
\end{eqnarray}
$g_2$ and $g_Y$ are the gauge coupling constants of SU(2)$_L$ and
U(1)$_Y$ gauge interactions.
The lifetime of $S$ is then
\begin{eqnarray}
 \tau_S = 5 \times 10^{-5}~{\rm sec} \times
N^{-2}
\left( \frac{m_S}{400~{\rm GeV}} \right)^{-11}
\left( \frac{m_{3/2}}{500~{\rm MeV}} \right)^{8}\ .
\end{eqnarray}

With this relatively long lifetime and the large initial amplitude of
$S$, it is reasonable that the coherent oscillation of $S$ dominates
over the energy density of the universe before $S$ starts to
decay. Although it is possible that $S$ domination does not happen by
assuming a presence of another long-lived matter density (such as the
inflaton oscillation), we do not consider such a case here.

The decay of $S$ produces radiation and it reheats the universe.
The reheating temperature after the $S$ decay is calculated to be
\begin{eqnarray}
 T_d \simeq 0.5 \times \sqrt{ \Gamma_H M_{\rm Pl}}
= 90~{\rm MeV} \times
N
\left( \frac{m_S}{400~{\rm GeV}} \right)^{11/2}
\left( \frac{m_{3/2}}{500~{\rm MeV}} \right)^{-4}\ .
\end{eqnarray}
This temperature needs to be larger than $2$~MeV in order for
standard BBN to happen~\cite{Ichikawa:2005vw}.
The baryon asymmetry and the dark matter component of the universe also
needs to be generated before BBN.
Since primordial radiation and matter are significantly diluted by the
entropy production from the $S$ decays, it is non-trivial whether we can
consistently obtain those two components.
We explain here the dark matter (gravitino) production from the $S$
decay and we will discuss a possible mechanism for baryogenesis later.

Under the assumption that the $S$-domination happens, the number density
of the gravitinos, $n_{3/2}$, from $S$ decays is given by
\begin{eqnarray}
 \frac{n_{3/2}}{s} = \frac{3}{4} \frac{T_d}{m_S} B_{3/2} \times 2\ ,
\label{eq:density}
\end{eqnarray}
where $B_{3/2}$ is the branching fraction into two gravitinos and $s$ is
the entropy density of the universe. 
Note here that the number density can be calculated only with the two
parameters in the Lagrangian, $m_S$ and $m_{3/2}$.
The partial decay width is
calculated from Eq.~(\ref{eq:gravitino}):
\begin{eqnarray}
 \Gamma_{3/2} = \frac{1}{96 \pi} \frac{m_S^3}{M_{\rm Pl}^2} 
\left( \frac{m_S}{m_{3/2}} \right)^2\ .
\end{eqnarray}
Therefore, the branching fraction $B_{3/2} ( = \Gamma_{3/2} / \Gamma_H)$
is given by
\begin{eqnarray}
 B_{3/2} = 2 \times 10^{-6} \times
N^{-2} 
\left( \frac{m_S}{400~{\rm GeV}} \right)^{-6}
\left( \frac{m_{3/2}}{500~{\rm MeV}} \right)^{6}\ .
\end{eqnarray}
The energy density is then
\begin{eqnarray}
 \Omega_{3/2} h^2 = 0.09 \times
N^{-1}
\left( \frac{m_S}{400~{\rm GeV}} \right)^{-3/2}
\left( \frac{m_{3/2}}{500~{\rm MeV}} \right)^{3}\ .
\end{eqnarray}
Comparing with the observed dark matter density of the universe,
$\Omega_{\rm CDM} h^2 = 0.10 \pm 0.02$
(2$\sigma$)~\cite{Spergel:2003cb}, we find that the gravitino is a
perfect candidate of dark matter in this scenario.

If $S$ can decay into two Binos and if the Bino is the next to lightest
supersymmetric particle (NLSP), the produced Binos later decay into
gravitinos. This possibility is excluded by the BBN constraints on the
Bino decay~\cite{Kawasaki:2004qu}.
The conclusion is the same for the case where the Higgsino is lighter
than the Bino.
This justifies the assumption that $S$ mainly decays into two Higgs
bosons. Other SU(2)$_L$ or SU(3)$_C$ charged supersymmetric particles
are heavier than the Bino in gauge mediation.

In fact, if the scalar tau lepton (stau) is lighter than the Bino, which
is possible if there is a large $A_{\tilde \tau}$-term from the direct
coupling between the Higgs fields and $S$ in the K{\" a}hler potential,
a small portion of the $m_S > 2 m_{\tilde{B}}$ region becomes viable.
Staus can reduce the number density by the pair annihilation process
after the production from $S$ decays.
The energy density of the stau before its decay is approximately
\begin{eqnarray}
 \Omega_{\tilde{\tau} } h^2 
\sim \Omega_{\tilde \tau}^{\rm th} h^2 \frac{T_f}{T_d}\ ,
\end{eqnarray}
where $T_f$ is the freezing out temperature, $T_f \sim m_{\tilde
\tau}/20$, and $\Omega_{\rm NLSP}^{\rm th} h^2 (\sim 0.002 (m_{\tilde
\tau}/100~{\rm GeV})^2)$ is the energy density of the stau obtained by
the standard calculation of the thermal relic abundance.
In the region of our interest, where the gravitinos from the $S$ decay
explain dark matter of the universe, the above amount is on the border
of the BBN constraints.
Since the viable region appears only in a special circumstance and also
the region is quite small, we do not consider this possibility in the
rest of the paper.

\vspace{5mm}
\noindent {\bf Case (B)}: The $S \to hh$ decay is closed ($m_S < 2 m_h$).\\ 
In this case, the dominant decay channel is into two gluons, $S \to
gg$. The decay rate is calculated to be
\begin{eqnarray}
 \Gamma_g = \frac{x_g^2 N^2}{96 \pi}
\frac{m_S^3}{M_{\rm Pl}^2}
\left( \frac{m_S}{m_{3/2}} \right)^4\ ,
\end{eqnarray}
with a one-loop factor
\begin{eqnarray}
 x_g \equiv \frac{g_s^2}{(4 \pi)^2} = 9.4 \times 10^{-3}\ .
\end{eqnarray}
 From this, the reheating temperature after the $S$ decay is given by
\begin{eqnarray}
 T_d \simeq 2~{\rm MeV} \times
N 
\left( \frac{m_S}{70~{\rm GeV}} \right)^{7/2}
\left( \frac{m_{3/2}}{15~{\rm MeV}} \right)^{-2}\ .
\end{eqnarray}

The branching fraction of the gravitino mode is 
\begin{eqnarray}
 B_{3/2} = 5 \times 10^{-4} \times 
N^{-2}
\left( \frac{m_S}{70~{\rm GeV}} \right)^{-2}
\left( \frac{m_{3/2}}{15~{\rm MeV}} \right)^{2}\ .
\end{eqnarray}
By using the same formula in Eq.~(\ref{eq:density}), we obtain
\begin{eqnarray}
 \Omega_{3/2} h^2 =
0.1 \times N^{-1}
\left( \frac{m_S}{70~{\rm GeV}} \right)^{1/2}
\left( \frac{m_{3/2}}{15~{\rm MeV}} \right)\ .
\end{eqnarray}
The dark matter density can be explained by the non-thermally produced
gravitinos also in this case.

\begin{figure}[t]
 \begin{center}
  \includegraphics[height=7.5cm]{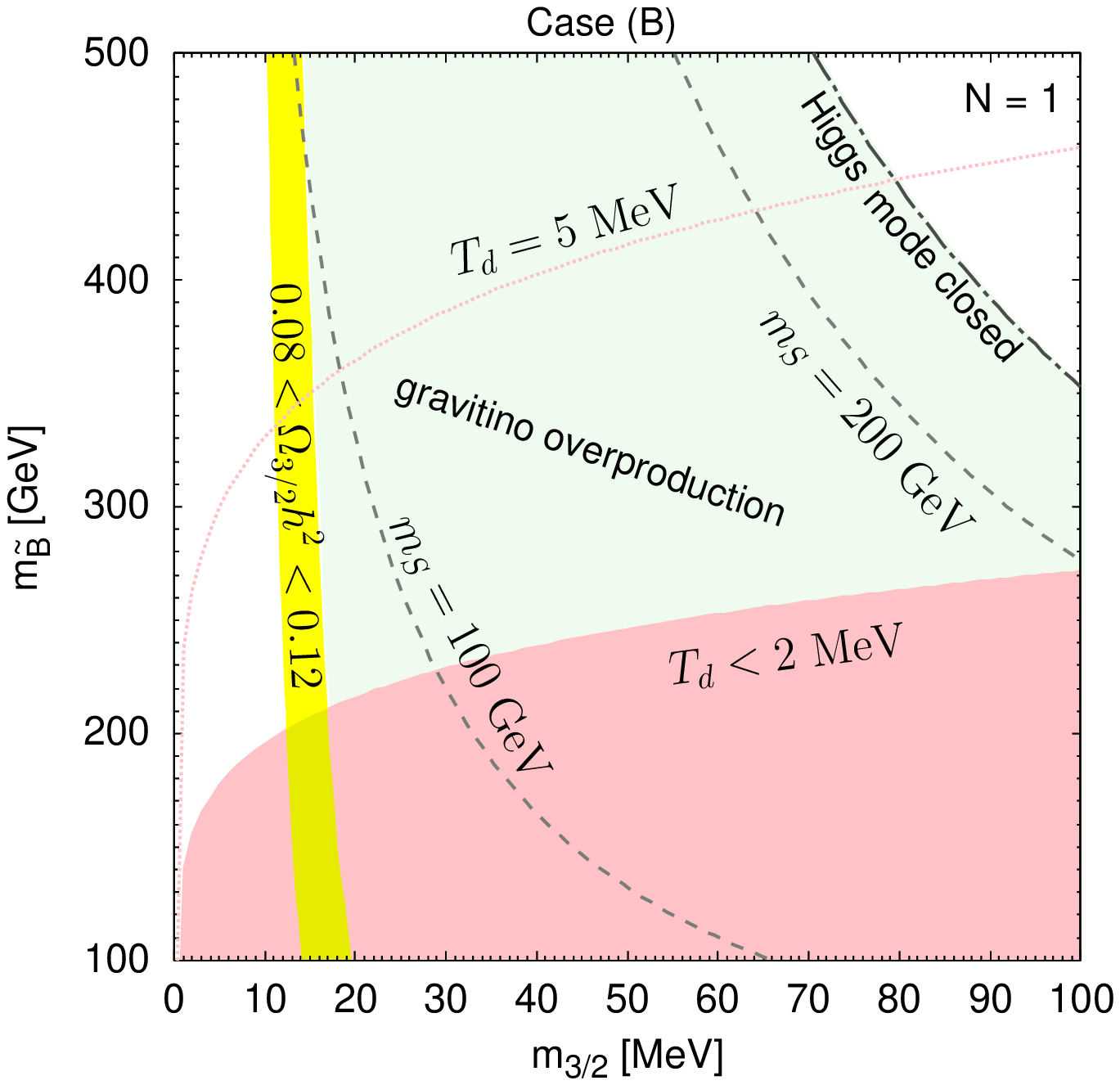}
  \includegraphics[height=7.5cm]{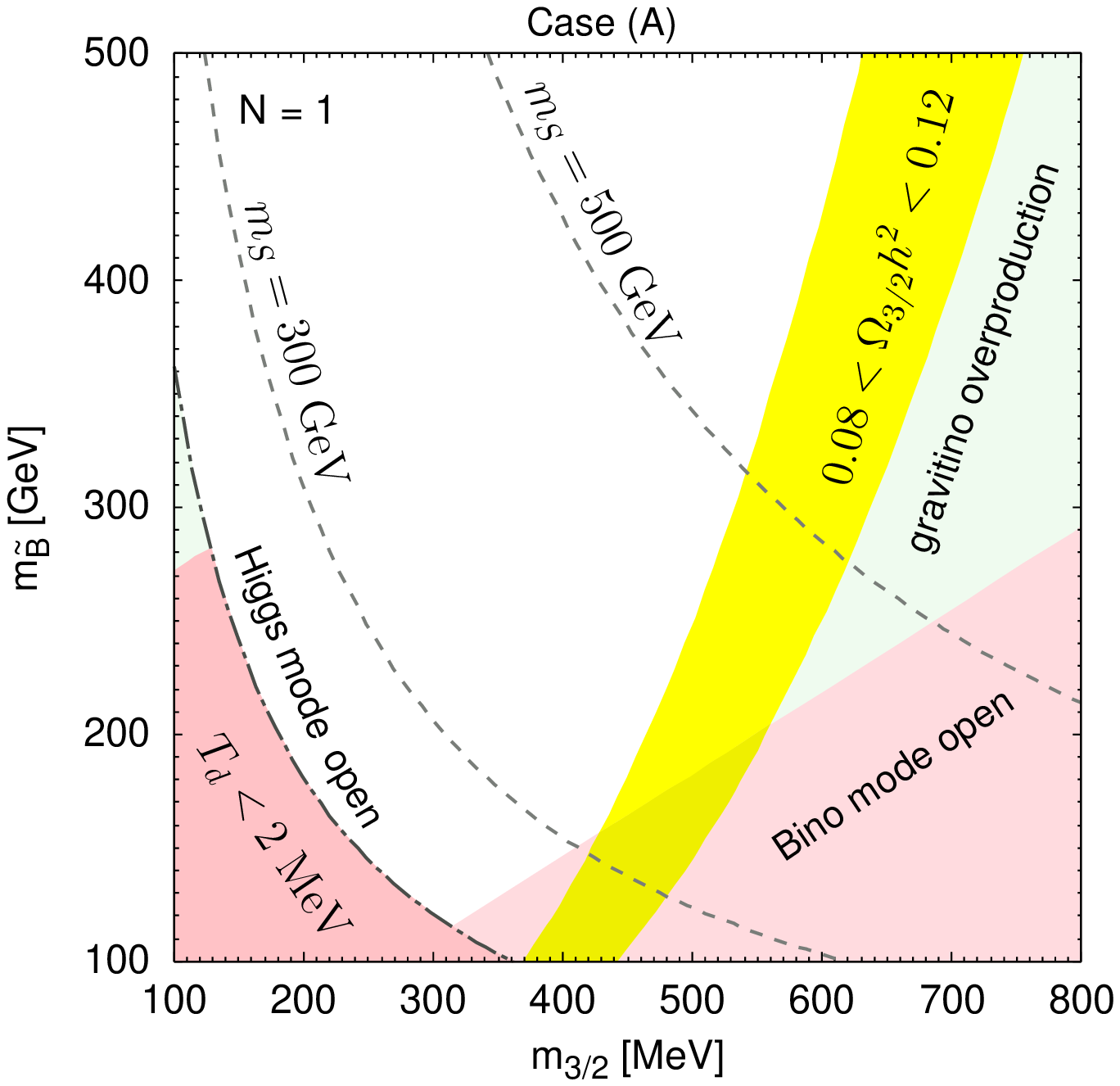}
\end{center}
\caption{
 The parameter regions of the model ($N=1$) which account for the dark
 matter component of the universe. The correct amount of gravitino dark
 matter can be obtained within the allowed range of parameter
 space. Below the dot-dashed line in the left figure, the decay mode of
 $S$ into two Higgs boson is closed (Case (B)). The constraint from the
 lower bound on the decay temperature $T_d > 2$~MeV is indicated. The
 contour of $T_d = 5$~MeV is also shown.
 In the right figure, the decay mode $S \to hh$ is open above the
 dot-dashed line (Case (A)). The decay mode into two Binos is open in
 the right-lower region.
}
\label{fig:region}
\end{figure}

\vspace{5mm}
We show in Fig.~\ref{fig:region} the region of the parameter space where
the non-thermally produced gravitino explains the dark matter component
of the universe.
We took $N=1$ and used $m_{3/2}$ and $m_{\tilde B}$ as free parameters
of this two-parameter model, and
showed contours of $m_S$ (see Eq.~(\ref{eq:bino})) as dashed lines.
The values of $\Omega_{3/2} h^2$ expressed in terms of those two are
\begin{eqnarray}
 \Omega_{3/2} h^2 
=
\left \{
\begin{array}{ll}
\displaystyle
0.09 \times N^{-1/4}
\left( \frac{m_{3/2}}{500~{\rm MeV}} \right)^{9/4}
\left( \frac{m_{\tilde B}}{200~{\rm GeV}} \right)^{-3/4}  & 
({\mbox{Case (A)}})\ , \vspace*{2mm}\\
\displaystyle
0.1 \times N^{-5/4}
\left( \frac{m_{3/2}}{15~{\rm MeV}} \right)^{5/4}
\left( \frac{m_{\tilde B}}{200~{\rm GeV}} \right)^{1/4}  & 
({\mbox{Case (B)}})\ ,\\
\end{array}
\right.
\end{eqnarray}
respectively.
The right figure corresponds to the Case (A) where the Higgs boson mode
is the dominant decay channel (above the dot-dashed line). The
constraint from $T_d > 2$~MeV is satisfied in the figure once the Higgs
mode is open.  The lower bound on the Bino mass ($m_{\tilde B} \gtrsim
160$~GeV) is obtained from the requirement that the Bino mode should be
kinematically forbidden. The gravitinos can explain the dark matter
density when $m_S \sim 300-500$~GeV, which is exactly the preferred
values from the $\mu$-term driven supersymmetry breaking scenario and
grand unified theories.

Another consistent region is found in the left figure, where the gluon
mode is the dominant decay channel (Case (B)).  The constraint from BBN,
$T_d > 2$~MeV, excludes the lower mass region of $m_{\tilde B}$ in this
case.
The correct amount of dark matter can be obtained with $m_S \sim
100$~GeV. This is also a natural size for the $\mu$-term driven
supersymmetry breaking scenario.

For a larger value of $N$, we obtain a larger lower bound on $m_{\tilde
B}$, and the allowed region of $m_{3/2}$ is not very sensitive to $N$
for Case (A) and moves to larger values linearly for Case (B).

We comment on other possible sources of gravitinos. Thermally produced
gravitinos~\cite{Moroi:1993mb,deGouvea:1997tn,Bolz:2000fu,Pradler:2006qh}
are significantly diluted by the entropy production from $S$ decays. It
is diluted by a factor of $T_d / T_{\rm dom}$ where $T_{\rm dom}$ the
temperature at which $S$ oscillation starts to dominate (see discussion
in the Subsection~\ref{sec:entropy}).
If $S$ oscillation starts during the reheating process after inflation,
which is the case in our scenario in order not to thermalize the
messenger fields, the amount of thermal gravitinos does not depend on
the reheating temperature after inflation. It is given in terms of the
parameters of the model and the initial amplitude $|S_0|$ by
\begin{eqnarray}
 \Omega^{\rm th}_{3/2} h^2
= \left \{
\begin{array}{ll}
\displaystyle
 8 \times 10^{-5}
\times N^{-11/4}
\left(
\frac{m_{3/2}}{500~{\rm MeV}}
\right)^{-13/4}
\left(
\frac{m_{\tilde B}}{200~{\rm GeV}}
\right)^{23/4}
\left(
\frac{|S_0|}{\Lambda}
\right)^{-2}
& \mbox{(Case (A))}\ ,\vspace*{2mm}\\ 
\displaystyle
 2 \times 10^{-3}
\times N^{-7/4}
\left(
\frac{m_{3/2}}{15~{\rm MeV}}
\right)^{-9/4}
\left(
\frac{m_{\tilde B}}{200~{\rm GeV}}
\right)^{19/4}
\left(
\frac{|S_0|}{\Lambda}
\right)^{-2}
& \mbox{(Case (B))}\ .\\
\end{array}
\right.
\label{eq:thermal}
\end{eqnarray}
This is sub-dominant when $|S_0| \sim \Lambda$.

Thermal relic of NLSPs also generates gravitinos through their
decays~\cite{Feng:2003xh}. Again entropy production reduces the amount
of NLSPs by a factor of ${\rm Max} [(T_d/T_f)^3, T_d / (T_{\rm dom}
T_f)^{1/2}, T_d / T_{\rm dom}]$, where $T_f (\sim m_{\rm NLSP}/20)$ is
the freezing out temperature of pair annihilation
processes~\cite{Buchmuller:2006tt}. The superWIMP component of gravitino
dark matter is therefore negligible in the parameter regions of our
interest.

\subsection{Are gravitinos cold?}

Since gravitinos are non-thermally produced through the decay of a heavy
particle $S$, they are relativistic when produced, but they subsequently
slow down as the universe expands.
The gravitinos need to be sufficiently non-relativistic before the
structure formation to start.
The free-streaming scale of the dark matter is constrained to be less
than $O(1)$~Mpc by the matter power spectrum inferred from
Lyman-$\alpha$ forest~\cite{Narayanan:2000tp}.

We estimate the free-streaming scale $ \lambda_{\rm FS}$ of the
gravitinos. We find that it is sufficiently small, but interestingly it
may be large enough to be distinguished from that of cold dark matter.
The free-streaming scale at the time of matter-radiation equality
$t_{\rm eq} $ is defined by
\begin{eqnarray}
 \lambda_{\rm FS} \equiv \int_{0}^{t_{\rm eq}} \frac{v(t)}{a(t)} dt
&\simeq& 
\frac{m_S}{2 m_{3/2}} \frac{a_D}{a_{\rm eq}^{1/2} H_0}
\left[
\log 
\frac{ 4 m_{3/2} }{ m_S }
\frac{ a_{\rm eq} }{ a_{D} }
\right]
\ ,
\end{eqnarray}
where $a_{\rm eq}$ and $a_D$ are scale factors at matter-radiation
equality and at the time of $S$ decays, respectively.  $H_0$ is the
Hubble parameter at the current epoch.
Up to logarithmic dependence, we obtain
\begin{eqnarray}
\lambda_{\rm FS} \simeq 
\left \{
\begin{array}{ll}
\displaystyle
0.004~{\rm Mpc} \times
N^{-1}
\left( \frac{m_S}{400~{\rm GeV}} \right)^{-9/2}
\left( \frac{m_{3/2}}{500~{\rm MeV}} \right)^{3} & \mbox{(Case (A))}\ ,
\vspace{2mm}\\
\displaystyle
0.5~{\rm Mpc} \times
N^{-1}
\left( \frac{m_S}{70~{\rm GeV}} \right)^{-5/2}
\left( \frac{m_{3/2}}{15~{\rm MeV}} \right) & \mbox{(Case (B))}\ .\\
\end{array}
 \right.
\end{eqnarray}

For Case (B), $\lambda_{\rm FS}$ is in an interesting range for the
small-scale structure problem in $\Lambda$CDM model.
Although not conclusive, it has been reported that there are
discrepancies between simulations of structure formation in the cold
dark matter scenario and observation at a small scale. The observed
number of galactic satellites in the inner region ($O({\rm Mpc})$) of
halo is an order of magnitude smaller than results of
simulations~\cite{Klypin:1999uc}. Also, simulation of the structure
formation predicts a cuspy profile in the central region of
halo~\cite{Moore:1997sg} whereas observations such as studies of
gravitational lensing~\cite{Tyson:1998vp} and rotation
curves~\cite{Simon:2004sr} indicate nearly uniform density core
structure. It has been studied that dark matter with free-streaming
scales of $O(0.1 - 1)$~Mpc may resolve these issues because it can smear
out the density fluctuation in small scale~\cite{Hogan:2000bv,
Kaplinghat:2005sy}. (See also \cite{Kitano:2005ge} for recent proposals
of particle physics models which realize large $\lambda_{\rm FS}$ via
non-thermally produced dark matter.)

A free-streaming scale of $O(10-100)$~kpc means that there is a cut-off
in matter power spectrum below that scale. It might be possible to probe
this small-scale structure in future experiments by using strong
gravitational lensing~\cite{Dalal:2002su} such as in LSST~\cite{LSST}
and SNAP~\cite{SNAP}, and also by observations of 21cm fluctuations in
high red-shift region~\cite{Loeb:2003ya}.

\subsection{Entropy production and baryogenesis}
\label{sec:entropy}

In the previous section, we have assumed that the oscillation energy
density dominates over the universe before it decays. We discuss the
amount of entropy production and possible baryogenesis scenarios here.

As we have discussed, the $S$ field starts to oscillate when the Hubble
parameter becomes smaller than $m_{S}$.
There are two cases which we need to consider separately.
If the oscillation starts during the reheating process after inflation
(the inflaton oscillation era), $S$-domination happens when:
\begin{eqnarray}
\label{eq:Tdom}
T_{\rm dom}\simeq
T_R\times \bigg(\frac{|S_{0}|}{\sqrt{3}M_{\rm Pl}}\bigg)^{2},
\end{eqnarray}
where $T_{R}$ is the reheating temperature after inflation.

A different result is obtained if $S$ starts to oscillate in a
radiation-dominated era (after reheating). The temperature at which $S$
starts to oscillate is given by
\begin{eqnarray}
\label{eq:oscillate}
 T_{\rm osc}\simeq 0.3 \times \sqrt{M_{\rm Pl} m_{S}} \simeq  8\times 10^{9}\, {\rm GeV }
\times \left(\frac{m_{S}}{400\,{\rm GeV}}\right)^{1/2}.
\end{eqnarray}
In this case, $T_{\rm dom}$ is
\begin{eqnarray}
\label{eq:Tdom2}
 T_{\rm dom} \simeq
T_{\rm osc}\times \bigg(\frac{|S_{0}|}{\sqrt{3}M_{\rm Pl}}\bigg)^{2}.
\end{eqnarray}
The dilution factor $\Delta^{-1}$ is then
\begin{eqnarray}
  \label{eq:dilution}
 \Delta^{-1} 
 \simeq\frac{T_d}{T_{\rm dom}} \simeq
  \left\{
   \begin{array}{lll}
   &\displaystyle{\frac{T_d}{T_R}}
 \bigg(\frac{|S_{0}|}{\sqrt{3}M_{\rm Pl}}\bigg)^{-2},
     \,\,(T_R<T_{\rm osc}),\vspace{2mm}\\
   & \displaystyle{\frac{T_d}{T_{\rm osc}}}
 \bigg(\frac{|S_{0}|}{\sqrt{3}M_{\rm Pl}}\bigg)^{-2},
     \,\,(T_R>T_{\rm osc}).
   \end{array}
  \right.
\end{eqnarray}
The primordial radiation and matter are diluted by this
amount. ($S$-domination does not happen if $\Delta^{-1}$ is larger than
unity.)
With the low decay temperature $T_d$, dilution effects are quite
large. For example, if $T_R < T_{\rm osc}$ and $S_0 \sim \Lambda \sim
M_{\rm GUT}$, the dilution factor is $\Delta^{-1} \sim 10^{-4} (T_R /
10^8~{\rm GeV})^{-1}$.
This dilutes unwanted relics such as thermally produced neutralinos to a
negligible level.
Since baryon asymmetry is also diluted, primordial baryon asymmetry
needs to be larger such as $n_B/s \sim 10^{-6}$.
It is a non-trivial task to obtain such a large asymmetry under a
condition that the reheating temperature cannot be too high in order not
to thermally produce the messenger particles (min$[ (T_R T_{\rm
osc})^{1/2},T_{\rm osc}] \ll \lambda \langle S \rangle$).

Examples of possible scenarios for baryogenesis are the right-handed
sneutrino inflation scenario~\cite{Murayama:1992ua} and scenario with
right-handed sneutrino dominated universe~\cite{Murayama:1993em}.
In those scenarios, baryon asymmetry is generated via
leptogenesis~\cite{Fukugita:1986hr} after the decay of right-handed
sneutrinos as follows:
\begin{eqnarray}
\label{eq:Baryonasymm}
 \frac{n_B}{s}
 \simeq
(0.2-0.8) \times10^{-8}
\times
\bigg(\frac{T_R}{10^{8}~{\rm GeV}}\bigg)
\bigg(\frac{m_{\nu 3}}{0.05~{\rm eV}}\bigg)
\sin\delta_{\rm eff},
\end{eqnarray}
where $\delta_{\rm eff}$ is an effective CP violating phase and $m_{\nu
3}$ corresponds to the heaviest neutrino mass, and $T_{R}$ denotes the
decay temperature of the right handed sneutrino.  After dilution by the
$S$ decay, it reduces to
\begin{eqnarray}
\frac{n_B}{s}
 \simeq
(0.2-1)\times 10^{-10}
\times
\left(\frac{T_{d}}{100\,\rm MeV} \right)
\left( \frac{|S_{0}|}{10^{15}\, \rm GeV} \right)^{-2}
\bigg(\frac{m_{\nu 3}}{0.05~{\rm eV}}\bigg)
\sin\delta_{\rm eff},
\end{eqnarray}
independent of $T_R$ for $T_{\rm osc}>T_{R}$.  Therefore, the observed
baryon asymmetry $n_{B}/s \simeq (8.7\pm0.3) \times
10^{-11}$~\cite{Spergel:2003cb} can be consistently explained with a
slightly suppressed initial amplitude $|S_{0}|$.
Note here that if the initial amplitude $S_0$ is suppressed to this
level, thermally produced gravitinos can give a non-negligible
contribution to the matter density (see Eq.~(\ref{eq:thermal})), and
also the decay of thermally produced NLSPs during BBN may put
constraints on species of the NLSP and/or $T_R$.

\section{Summary}
\label{sec:summary}

Taking into account gravity effects in supersymmetry breaking/mediation
models opens up a possibility of very simple gauge mediation. We have
considered the simplest model of gauge mediation, in which we have only
two parameters.
One combination of the parameters needs to be fixed so that we obtain
the gaugino masses to be $O(100)$~GeV, and the other has preferable
range from a simple solution to the $\mu$-problem and grand unification,
i.e., 10~MeV $\lesssim m_{3/2} \lesssim 1$~GeV.

This is the simplest model of gauge mediation, but amazingly free from
problems.
We have shown that $S$ rolls into the supersymmetry breaking vacuum with
generic initial conditions. The entropy production by the $S$ decay
dilutes gravitinos and neutralinos which are thermally (over)produced.
Supersymmetric particles do not cause the FCNC or CP problems
as the gravitino mass is small enough.
Of the most amazing is the consistency of the dark-matter abundance. We
can calculate the abundance of the non-thermally produced gravitinos
from $S$ decays in terms of the two parameters in the model.
Within the parameter region of our interest, the calculated amount can
account for the dark matter of the universe.

\section*{Acknowledgments}

We thank Ted Baltz and Masamune Oguri for useful discussion on dark
matter. R.K. would like to thank Paddy Fox, Hitoshi Murayama and
Yasunori Nomura for stimulating discussion and Jay Wacker for discussion
on non-perturbative particle production during the $S$ oscillation.
This work was supported by the U.S. Department of Energy under contract
number DE-AC02-76SF00515.

\appendix 
\section{Initial condition of {\boldmath $S$} }
\label{app:amplitude}

As we mentioned in section~\ref{sec:evolution}, the minimum of the
scalar potential during inflation is likely to be around $|S|\sim
\Lambda$.

The singlet field $S$ may couple to the inflaton sector in the
superpotential,
\begin{eqnarray}
\label{eq:StoInf}
 W = W_{\rm Inflaton} \left(  1 - \xi\frac{\Lambda}{M_{\rm Pl}^{2}} S + \cdots \right),
 \end{eqnarray}
where $\xi$ is a coefficient of ${\cal O}(1)$, and $W_{\rm Inflaton}$
denotes the superpotential of a inflaton sector.  The scalar potential
of $S$ during inflation is then given by
\begin{eqnarray}
 \label{eq:HubbleMass}
 V(S) \simeq 
3 H^{2} \left( |S|^{2} -  \frac{|S|^{4}}{\Lambda^{4}} +\cdots \right)
+ 3 H^{2} \left| M_{\rm Pl} - \xi \frac{\Lambda}{M_{\rm Pl}} S \right|^{2},
\end{eqnarray}
where the first term originates from the K\"ahler potential in Eq.~(1)
and the second term from the superpotential in Eq.~(\ref{eq:StoInf}).
In the above expression, we have used the Hubble equation,
\begin{eqnarray*}
H^{2} = \frac{ |F_{\rm Inflaton}|^{2}}{3 M_{\rm Pl}^{2}}, 
\end{eqnarray*}
assuming that the $F$-component of the inflaton dominates the energy
density during inflation.  This potential has a minimum at $S_0= {\cal
O}(\Lambda)$.

In principle, the $\xi$ parameter can be much larger than $O(1)$
depending on a UV completion of the theory. For example, in a model of
Ref.~\cite{Kitano:2006wm}, the theory above the scale $\Lambda$ is a
strongly coupled conformal field theory, and the $\xi$ parameter is
enhanced to be $O(M_{\rm Pl} / \Lambda)$.
If this is the case, the second term in Eq.~(\ref{eq:HubbleMass}) pushes
the minimum to the Planck scale.
However, the effective theory in terms of $S$ does not make sense for
$|S|\gsim \Lambda$, and it is more appropriate to consider the dynamics
in terms of the constituent field $T$. 
The constituent field $T$, in contrast to $S$, may have a positive
Hubble mass squared around the symmetry enhancement point at $T=0$.
This Hubble mass term attracts $T$ to the origin, and it prevents $S\sim
TT$ from going to the Planck scale.  Therefore, we can expect the actual
minimum during inflation does not exceed ${\cal O}(\Lambda)$ even if the
coefficient $\xi$ is large.

If $\xi$ is large, higher order terms of $S$ in Eq.~(\ref{eq:StoInf})
can be also important.
Even in this case, the result of our discussion will not change
qualitatively.

\end{document}